\documentclass[aps,prl,twocolumn,floatfix,reprint]{revtex4}
\usepackage{amsfonts}
\usepackage{mathrsfs}
\usepackage{amsmath}
\usepackage{color}
\usepackage{graphicx}
\usepackage{bm}
\usepackage{amssymb}
\usepackage{xspace}
\usepackage{epstopdf}
\usepackage{dcolumn}
\usepackage{tabularx}
\usepackage{longtable}
\usepackage[colorlinks=true, pdfstartview=FitV, linkcolor=blue, citecolor=blue, urlcolor=blue]{hyperref}

\begin{document}

\title{Observation of Floquet band topology change in driven ultracold Fermi
gases}
\author{Lianghui Huang$^{1,2}$}
\author{Peng Peng$^{1,2}$}
\author{Donghao Li$^{1,2}$}
\author{Zengming Meng$^{1,2}$}
\author{Liangchao Chen$^{1,2}$}
\author{Chunlei Qu$^{3,4}$}
\author{Pengjun Wang$^{1,2}$}
\author{Chuanwei Zhang$^{3}$}
\email{chuanwei.zhang@utdallas.edu}
\author{Jing Zhang$^{1,5}$}
\email{jzhang74@sxu.edu.cn, jzhang74@yahoo.com}
\affiliation{$^{1}$State Key Laboratory of Quantum Optics and Quantum Optics Devices,
Institute of Opto-Electronics, Shanxi University, Taiyuan 030006, P.R.China }
\affiliation{$^{2}$Collaborative Innovation Center of Extreme Optics, Shanxi University,
Taiyuan 030006, P.R.China}
\affiliation{$^{3}$Department of Physics, The University of Texas at Dallas, Richardson,
Texas 75080, USA}
\affiliation{$^{4}$INO-CNR BEC Center and Dipartimento di Fisica, Universit\`a di Trento,
38123 Povo, Italy}
\affiliation{$^{5}$Synergetic Innovation Center of Quantum Information and Quantum
Physics, University of Science and Technology of China, Hefei, Anhui 230026,
P. R. China}

\begin{abstract}
Periodic driving of a quantum system can significantly alter its energy
bands and even change the band topology, opening a completely new avenue for
engineering novel quantum matter. Although important progress has been made
recently in measuring topological properties of Floquet bands in different
systems, direct experimental measurement of Floquet band dispersions and
their topology change is still demanding. Here we directly measure Floquet
band dispersions in a periodically driven spin-orbit coupled ultracold Fermi
gas. Using spin injection radio-frequency spectroscopy, we observe that the
Dirac point originating from two dimensional spin-orbit coupling can be
manipulated to emerge at the lowest or highest two dressed bands by fast
modulating Raman laser frequencies, demonstrating topological change of
Floquet bands. Our work will provide a powerful tool for understanding
fundamental Floquet physics as well as engineering exotic topological
quantum matter.
\end{abstract}

\maketitle

Engineering energy band dispersions plays a crucial role for designing
quantum materials with novel functionalities. Besides traditional methods in
solid state, periodic modulation of system parameters can significantly
alter the band dispersions of a quantum matter such as turning a trivial
insulator into a topological one~\cite{Linder2011}. Thanks to Floquet
theory, such periodic driven quantum systems can be described by an
effective static Floquet Hamiltonian, which may exhibit distinct properties
compared to their unmodulated counterparts. Experimentally, such Floquet
band engineering has been recently investigated in atomic~\cite%
{Aidelsburger2013,Miyake2013}, photonic~\cite{Rechtsman2013} and solid state
systems~\cite{Wang2013}.

Ultracold atomic gases, due to its unprecedented tunability, provides an
ideal platform for the investigation of Floquet physics~\cite{Eckard2016}.
As a prominent example, by loading ultracold atoms in a periodically
modulated optical honeycomb lattice, recent experiment \cite{Jotzu2014} has
realized the Haldane model that exhibits anomalous quantum Hall effect~\cite%
{Haldane1988}. So far, cold atom experiments have mainly focused on
detecting Floquet band structures and their properties indirectly, such as
through atomic transport \cite{Jotzu2014} or by adiabatically loading
bosonic atoms to band minima~\cite{Parker2013,Spielman2015,Khamehchi2016}. A
direct measurement of the Floquet band dispersions and the change of their
topological properties is still lacking in atomic systems.

Dirac points are band touching points with linear dispersions and their
creation and annihilation showcase one type of topological change of band
dispersions of a quantum matter. Two-dimensional (2D) spin-orbit coupling
(SOC), such as Rashba SOC, naturally possesses a Dirac point in its band
dispersion. It is well known that SOC plays a key role in many exotic
topological materials \cite{Hasan,Qi}. In ultra-cold atoms, synthetic
one-dimension (1D) SOC (an equal sum of Rashba \cite{Rashba} and Dresselhaus
\cite{Dresselhaus55PR} SOC) was first experimentally realized using a pair
of counter-propagating Raman lasers to dress two atomic spin states \cite%
{spielman,FuPRA,Shuai-PRL,Washington-PRA,Purdue,Jing,MIT,Spielman-Fermi,Lev,Jo}%
. Recently, by coupling three internal spin states of ultracold $^{40}$K
Fermi gases through three Raman lasers propagating in a plane, a 2D SOC
characterized by the emergence of a Dirac point has been observed~\cite%
{Huang15}. 
Furthermore, an energy gap which is crucial for the investigation of
topological physics in ultracold atomic gases can be generated at the
position of the Dirac point by tuning the polarization of the Raman lasers
\cite{Meng15}.

In this paper, we utilize 2D spin-orbit coupled Fermi gases as a platform to
investigate Floquet band engineering. By periodically modulating the
detunings of two Raman lasers through their frequencies, we can manipulate
the strengths and even the signs of the Raman coupling of the effective
Floquet Hamiltonian and therefore modify the position of the Dirac point in
the Floquet band. For suitable modulations, the Dirac point initially
located at the lower two dressed bands can disappear and then emerge at the
upper two bands. Such modulation induced Floquet bands and their topology
change are directly observed and characterized in experiment using spin
injection radio-frequency (rf) spectroscopy. The corresponding Floquet
sidebands are also observed in experiments. Our results showcase the 2D
spin-orbit coupled Fermi gas as a powerful platform for exploring Floquet
band engineering and exotic quantum matter~\cite%
{Goldman2014,Eckardt2015,Bukov2015}.

{\Large Results}

\begin{figure}[tbp]
\centerline{
\includegraphics[width=8cm]{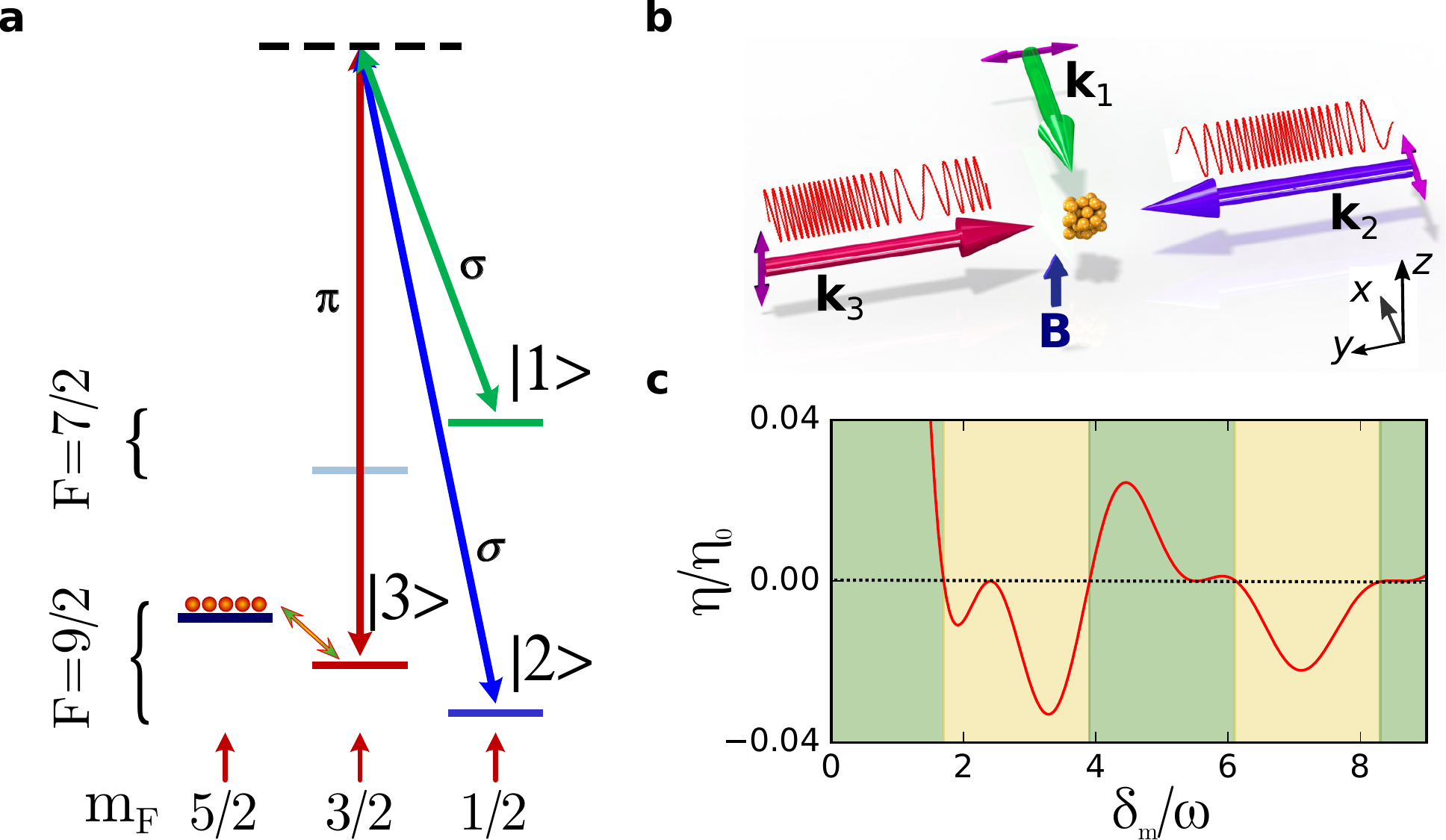}}
\caption{\textbf{Scheme for realizing driven degenerate Fermi gases with
modulated 2D SOC}. (a) Energy level diagram of Fermi gases $^{40}$K. Three
hyperfine spin states are coupled with the electronic excited states through
three Raman lasers. The atoms are initially prepared in the free reservoir
spin state $|9/2,5/2\rangle $. (b) Configuration of three Raman lasers in
the $xy$ plane. The detunings of the Raman lasers 2 and 3 are modulated as $%
\protect\delta _{2(0)}+\protect\delta _{m}\cos (\protect\omega t)$ and $%
\protect\delta _{3(0)}+\protect\delta _{m}\cos (\protect\omega t+\protect%
\phi _{0})$. (c) Plot of the product of the effective Raman coupling
strengths $\protect\eta =\Omega _{12}^{\prime }\Omega _{13}^{\prime }\Omega
_{23}^{\prime }$ (scaled by $\protect\eta _{0}=\Omega _{12}\Omega
_{13}\Omega _{23}$) as a function of the modulation parameter $\protect%
\delta _{m}/\protect\omega $. The background colors indicate the sign of $%
\protect\eta $ which determines the position of the Dirac point. The
relative phase $\protect\phi _{0}=\protect\pi /2$.}
\label{Fig1}
\end{figure}

The experimental setup for generating 2D SOC is the same as that in our
previous experiment~\cite{Huang15} (see \textbf{Methods}). As shown in Fig.~%
\ref{Fig1}(a), three hyperfine spin states of the $^{40}$K Fermi gas are
coupled to the electronic excited states by three far-detuned Raman lasers,
with the corresponding two-photon Raman coupling strengths between hyperfine
states $|j\rangle $ and $|j^{\prime }\rangle $ denoted by $\Omega
_{jj^{\prime }}$. The three Raman lasers propagate in the $x$-$y$ plane
(Fig.~\ref{Fig1}(b)), thus the motion of the atoms along $z$ direction is
decoupled from the internal degrees of freedom.

In experiment, the two-photon Raman detunings are modulated as $\delta
_{2}=\delta _{2(0)}+\delta _{m}\cos (\omega t)$ and $\delta _{3}=\delta
_{3(0)}+\delta _{m}\cos (\omega t+\phi _{0})$ by varying the frequencies of
the Raman lasers 2 and 3 (see supplementary information \cite{supp}). Here, $%
\phi _{0}$ is the initial relative phase between the two modulations, which
could be tuned arbitrarily in experiment. With a high modulation frequency $%
\omega =2\pi \times 100$ kHz which is much larger than the other relevant
energy scales, the system can be described by an effective static Floquet
Hamiltonian (see Eq. (\ref{H_eff}) in \textbf{Methods}), which is the same
as that of the unmodulated system with the Raman coupling strengths $\Omega
_{jj^{\prime }}$ replaced by $\Omega _{12}^{\prime }=\Omega _{12}J_{0}\left(
\delta _{m}/\omega \right) $, $\Omega _{13}^{\prime }=\Omega
_{13}J_{0}\left( \delta _{m}/\omega \right) $, and $\Omega _{23}^{\prime
}=\Omega _{23}J_{0}\left( 2\delta _{m}\sin (\phi _{0}/2)/\omega \right) $
\cite{supp}. Here $J_{0}(x)$ is the $0$-th order Bessel function. The
effective $3\times 3 $ static Floquet Hamiltonian (\ref{H_eff}) has three
dressed bands, and the position of the Dirac point is determined by the sign
of the quantity $\eta =\Omega _{12}^{\prime }\Omega _{13}^{\prime }\Omega
_{23}^{\prime }$. The Dirac point emerges at the crossing of the lower
(upper) two bands for negative (positive) $\eta $~\cite{Huang15}. By varying
the modulating amplitude $\delta _{m}$ and relative phase $\phi _{0}$, we
can manipulate $\eta $ (see Fig.~\ref{Fig1}(c)) and thus alter the topology
of the Floquet band structure. For the measurement, we use momentum-resolved
spin injection rf spectroscopy to study the energy-momentum dispersions of
the dressed states, in which the atoms are driven from a free spin-polarized
state (initial state) into the SOC dressed ones (finial states).

\begin{figure*}[t]
\centerline{
\includegraphics[width=1.0\textwidth]{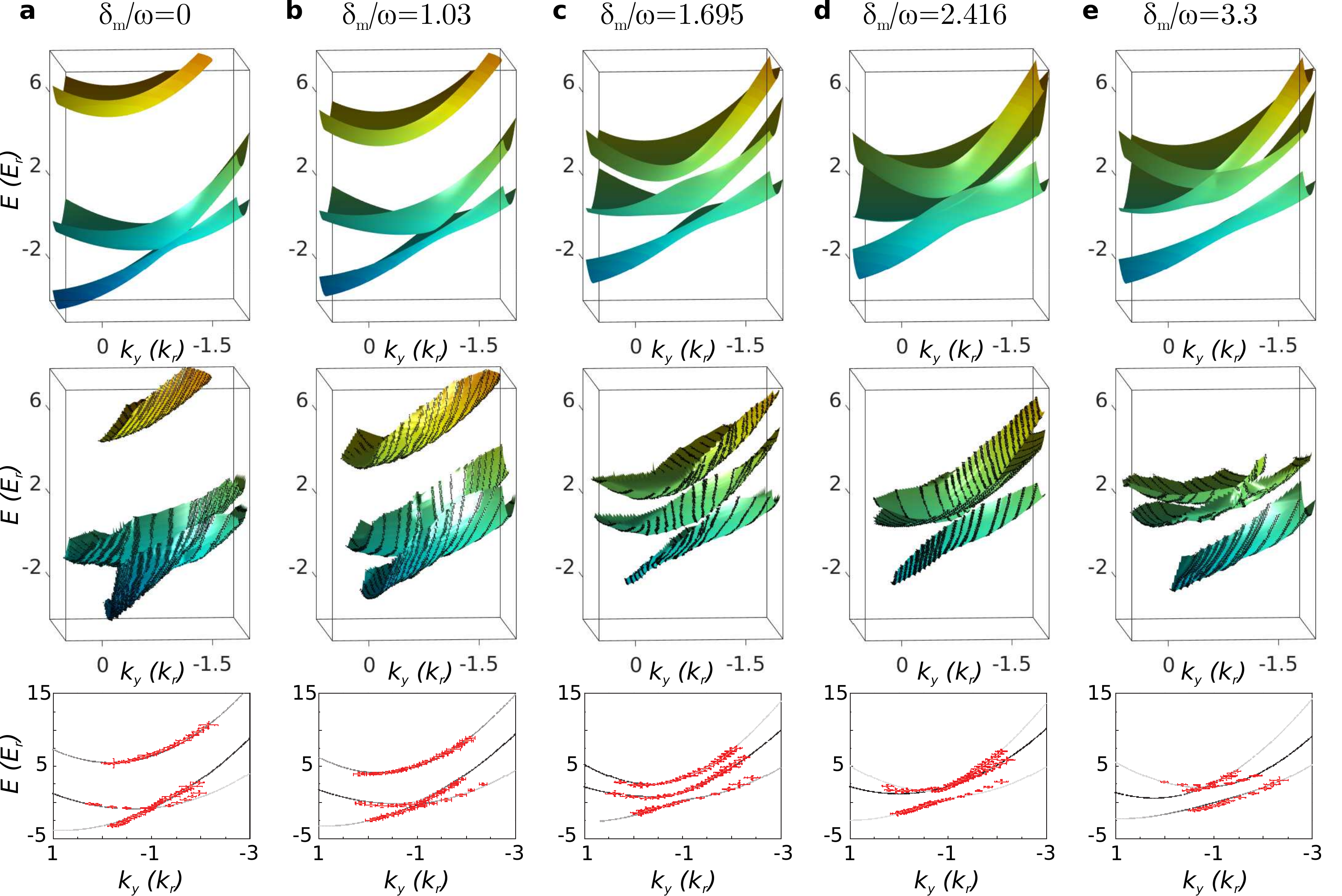}}
\caption{\textbf{Observation of Floquet band topological change}.
The upper and middle panels represent theoretically calculated and
experimentally measured 2D Floquet band dispersions, respectively.
The lower panel represents the corresponding 1D dispersions along
$k_{x}=0.05k_{r}$ for (a)-(d) and $k_{x}=-0.7k_{r}$ for (e) where
the greyness of the lines indicates the population of the final state $%
\left\vert 3\right\rangle $. The modulation amplitudes of the Raman
detunings are (a) $\protect\delta _{m}/\protect\omega =0$, (b) $1.03$, (c) $%
1.695$, (d) $2.416$, and (e) $3.3$. Other parameters are $\Omega
_{12}=5.46E_{r}$, $\Omega _{13}=4.62E_{r}$, $\Omega _{23}=-4.2E_{r}$, $%
\protect\delta _{2(0)}=-2.47E_{r}$, and $\protect\delta _{3(0)}=0.93E_{r}$.
The relative phase is $\protect\phi _{0}=\protect\pi /2$. }
\label{Fig2}
\end{figure*}

\emph{Observation of topology change of Floquet band dispersions}: The
wavelengths of the Raman lasers are tuned to 768.85 nm between the $D_{1}$
line and $D_{2}$ line, making $\eta =-1$ for the Raman coupling strengths in
the absence of the modulation $\delta _{m}=0$. Therefore, the two lower
energy bands touch at a Dirac point which is observed in experiment as shown
in Fig.~\ref{Fig2}(a) with the three band dispersions measured by
spin-injection rf spectroscopy. The band dispersions are also determined
theoretically by calculating the eigenenergy spectrum of the effective
static Floquet Hamiltonian~(\ref{H_eff}) and compared with the experimental
results.

We first consider the periodic modulations with the relative phase $\phi
_{0}=\pi /2$. By increasing $\delta _{m}/\omega $, the three effective Raman
coupling strengths decrease, and the Dirac point moves in the momentum
space, but still within the lowest two bands for a small $\delta _{m}/\omega
$ (see Fig. \ref{Fig2}(b)). When $\delta _{m}/\omega \approx 1.7$, $J_{0}(%
\sqrt{2}\delta _{m}/\omega )=0$ and thus $\Omega _{23}^{\prime }=0$. The two
spin states $|2\rangle $ and $|3\rangle $ decouple and the Dirac point moves
to infinity, showing three Floquet bands which are gapped everywhere (see
Fig. \ref{Fig2}(c)). When $\delta _{m}/\omega $ is slightly larger than $1.7$%
, the Dirac point re-appears at the crossing of the two upper bands because $%
\Omega _{23}^{\prime }$ changes sign and becomes positive. When $\delta
_{m}/\omega \approx 2.4$, $J_{0}(\delta _{m}/\omega )=0$ and thus $\Omega
_{12}^{\prime }=\Omega _{13}^{\prime }=0$, the two spin states $|2\rangle $
and $|3\rangle $ are coupled by an effective 1D SOC which does not exhibit a
Dirac point. For our experimental parameters, the uncoupled free particle
dispersion band for spin state $|1\rangle $ intersects with the upper branch
of the 1D SOC (Fig. \ref{Fig2}(d)), where a small gap between these two
dispersions is opened due to the finite driving frequency $\omega $. By
further increasing $\delta _{m}/\omega $, $\Omega _{12}^{\prime }$ and $%
\Omega _{13}^{\prime }$ change sign simultaneously and thus the Dirac point
remains staying at the same crossing of the two upper dressed bands (see
Fig. \ref{Fig2}(e)). Now the upper band initially without topological
properties becomes topological.

\begin{figure}[b]
\centerline{
\includegraphics[width=8.6cm]{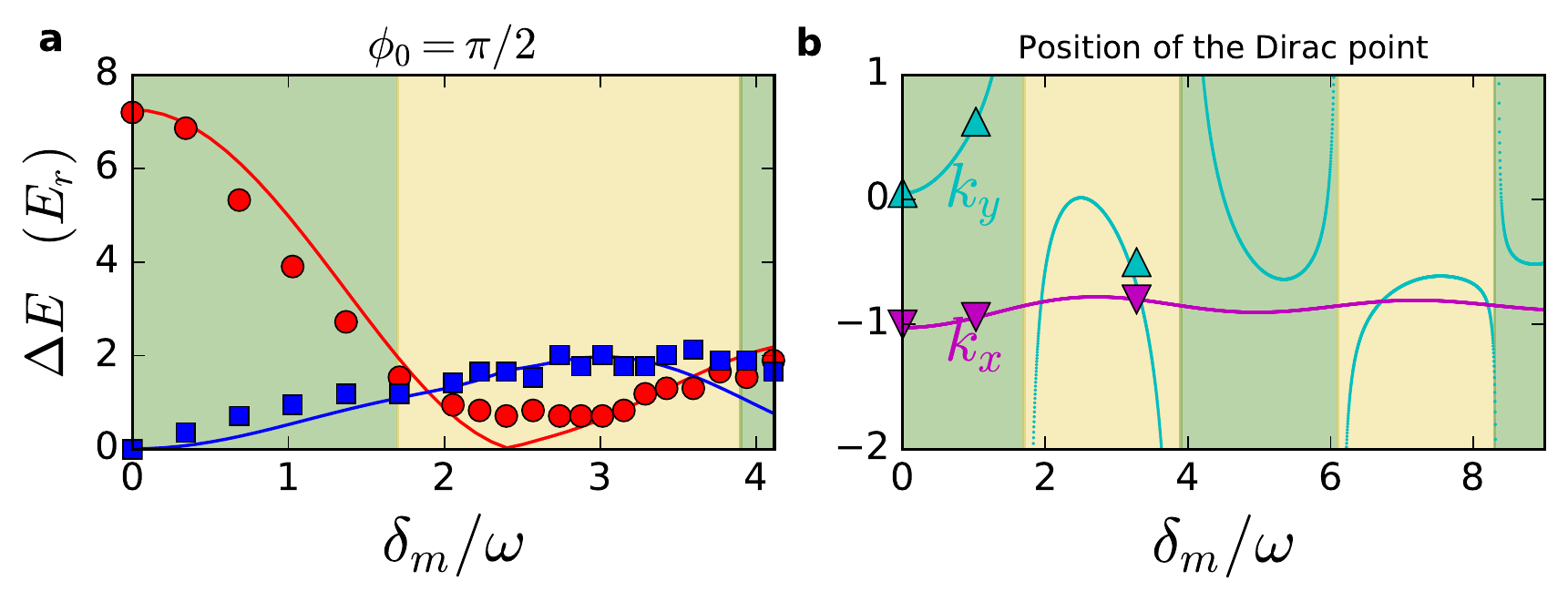}}
\caption{\textbf{Characterization of the change of the Floquet band structure%
}. (a) Energy separations between three energy bands at $\mathbf{k}_{0}$,
the position of the Dirac point in the absence of modulation. The symbols
and solid lines correspond to experimental and theoretical results,
respectively. (b) Change of the position of the Dirac point as a function of
the modulation parameter $\protect\delta _{m}/\protect\omega $. The magenta
and cyan lines correspond to theoretical plot of $k_{x}$ and $k_{y}$ of the
Dirac point. The symbols correspond to the positions of the measured Dirac
point for the three $\protect\delta _{m}/\protect\omega =0$, $1.03$, $3.3$.
Note that the Dirac point goes to infinity for $\protect\delta _{m}/\protect%
\omega =1.7$, as observed in Fig. \protect\ref{Fig2}(c). All other
parameters are the same as Fig.~\protect\ref{Fig2}. The background colors in
both panels indicate whether the Dirac point exhibits at the crossing of two
lower (green) or higher (yellow) bands.}
\label{Fig3}
\end{figure}

We denote $\mathbf{k}_{0}=(k_{x}^{0},k_{y}^{0})$ as the original position of
the Dirac point in momentum space in the absence of modulation. In the
presence of modulation, the position of the Dirac point is shifted to a
different place and there is an energy separation at $\mathbf{k}_{0}$
between the two crossed bands. We characterize the three band dispersions by
measuring the energy separations between the three dressed bands at the
position of $\mathbf{k}_{0}$. In Fig.~\ref{Fig3}(a), we plot these energy
separations as a function of the modulation parameter $\delta _{m}/\omega $.
With the increase of $\delta _{m}/\omega $, the three effective Raman
coupling strengths are decreased, the energy separation between the lower
two bands at $\mathbf{k}_{0}$ is increased (blue line in Fig.~\ref{Fig3}%
(a)), while the separation between the upper two bands is decreased (red
line in Fig.~\ref{Fig3}(a)). The good agreement between experiment and
theory demonstrate the expected modulation of the Floquet band dispersion.

In the presence of modulation, the current position of the Dirac point $%
\mathbf{k=}\left( k_{x},k_{y}\right) $ is different from $\mathbf{k}_{0}$.
For different values of $\delta _{m}/\omega $, it can be computed
theoretically from the effective static Floquet Hamiltonian (\ref{H_eff}).
In Fig.~\ref{Fig3}(b), we show the trajectory of the current Dirac point as
a function of the modulation parameter $\delta _{m}/\omega $, together with
the experimental measured positions for three values of $\delta _{m}/\omega $
shown in Fig.~\ref{Fig2}. Across the points $\delta _{m}/\omega =z_{n,0}/%
\sqrt{2}=1.7,3.9,6.1,8.3,\ldots $, the Dirac point moves to infinity (Fig. %
\ref{Fig2}(c)) and then reappears at the crossing of the other two dressed
bands (Fig. \ref{Fig2}(e)). Here $z_{n,0}$ are the zeros of the Bessel
function $J_{0}(z_{n,0})=0$. On the contrary, at the two sides of $\delta
_{m}/\omega =2.4,5.5,8.65,11.8,\ldots $, two of the effective Raman
couplings change sign simultaneously and the position of the Dirac point
does not change. Such observed move of the Dirac point between lower and
upper two bands with increasing $\delta _{m}/\omega $ showcases the topology
change of Floquet band structure of driven Fermi gases.

\begin{figure}[tbp]
\centerline{
\includegraphics[width=8.6cm]{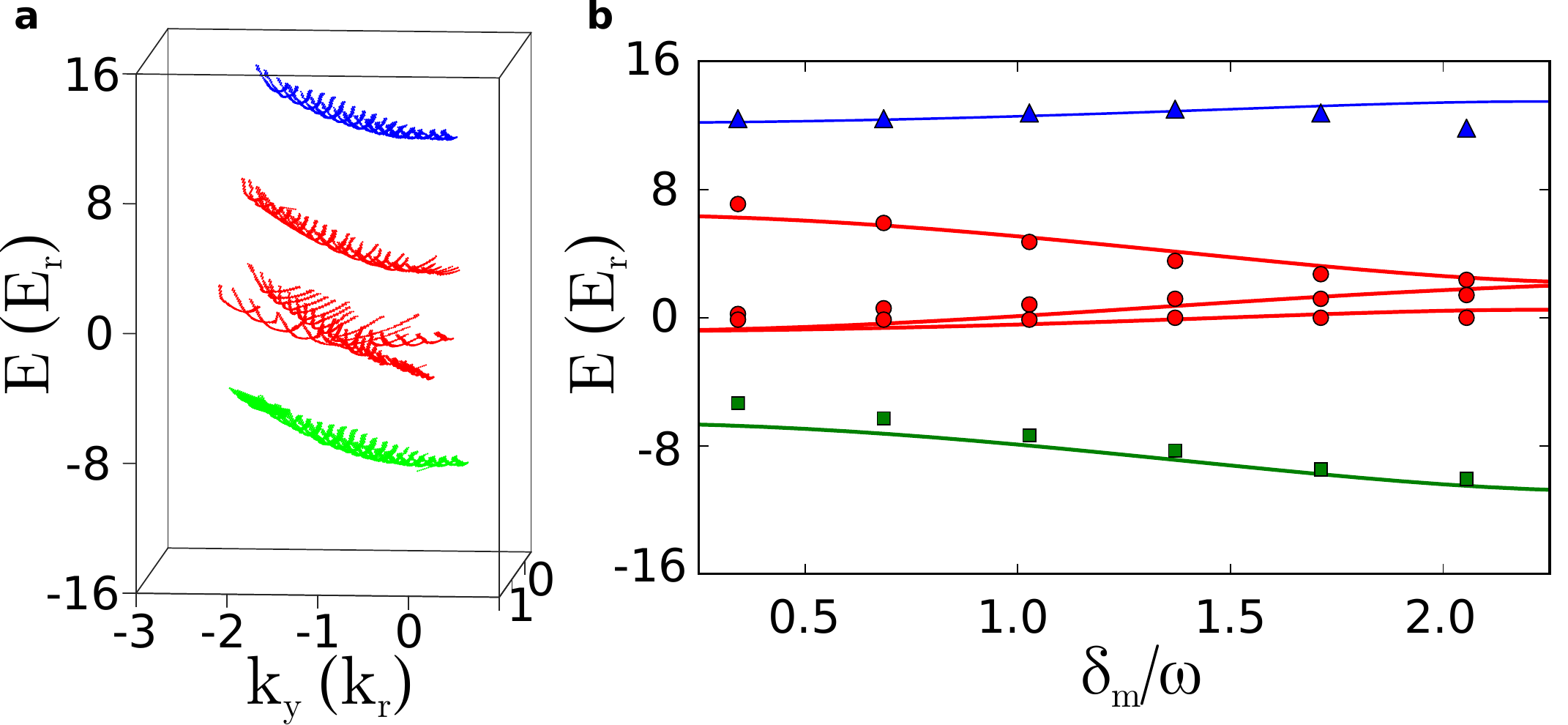}} 
\caption{\textbf{Observation of Floquet sidebands}. (a) Plot of the measured
energy bands by rf spin-injection spectroscopy with $\protect\delta _{m}/%
\protect\omega =1.03$. (b) Values of the quasi-energies at $\mathbf{k}_{0}$
as a function of $\protect\delta _{m}/\protect\omega $. The solid lines are
the theoretical plots and the symbols are the experimental data. The other
parameters are the same as those in Fig.~\protect\ref{Fig2}. }
\label{Fig4}
\end{figure}

\emph{Observation of Floquet sidebands}: Periodic driving not only modifies
the band structure drastically, but also induces Floquet sidebands. Due to
the absorption and emission of integer times of energy $\hbar \omega $, the
energy dispersion of the Floquet system repeats itself periodically in the
energy domain. Although the sidebands are not easy to be identified in
periodically driven bosonic systems, the modulated Fermi gas provides an
ideal platform to map out the sidebands using spin injection rf
spectroscopy. Fig.~\ref{Fig4}(a) shows the measured $n=+1$ and $n=-1$
sidebands. According to Floquet theory, the sideband energy dispersions
should be a simple copy of $n=0$ dressed energy bands. Here one dressed
energy band in $n=+1$ (or $n=-1$) sideband is shown in Fig.~\ref{Fig4}(a).
To test the Floquet theory, we also measure the energy separation of the
dressed states at the position of the original Dirac point $\mathbf{k}_{0}$
by including the $n=+1$ and $n=-1$ sidebands as a function of the modulation
parameter $\delta _{m}/\omega $. They agree with the theoretical calculation
very well as shown in Fig. \ref{Fig4}(b). The energy separation between the
sidebands and the original bands is around $\sim \hbar \omega =h\times 100$
kHz, the same as the driving frequency.

\begin{figure}[b]
\centerline{
\includegraphics[width=8.6cm]{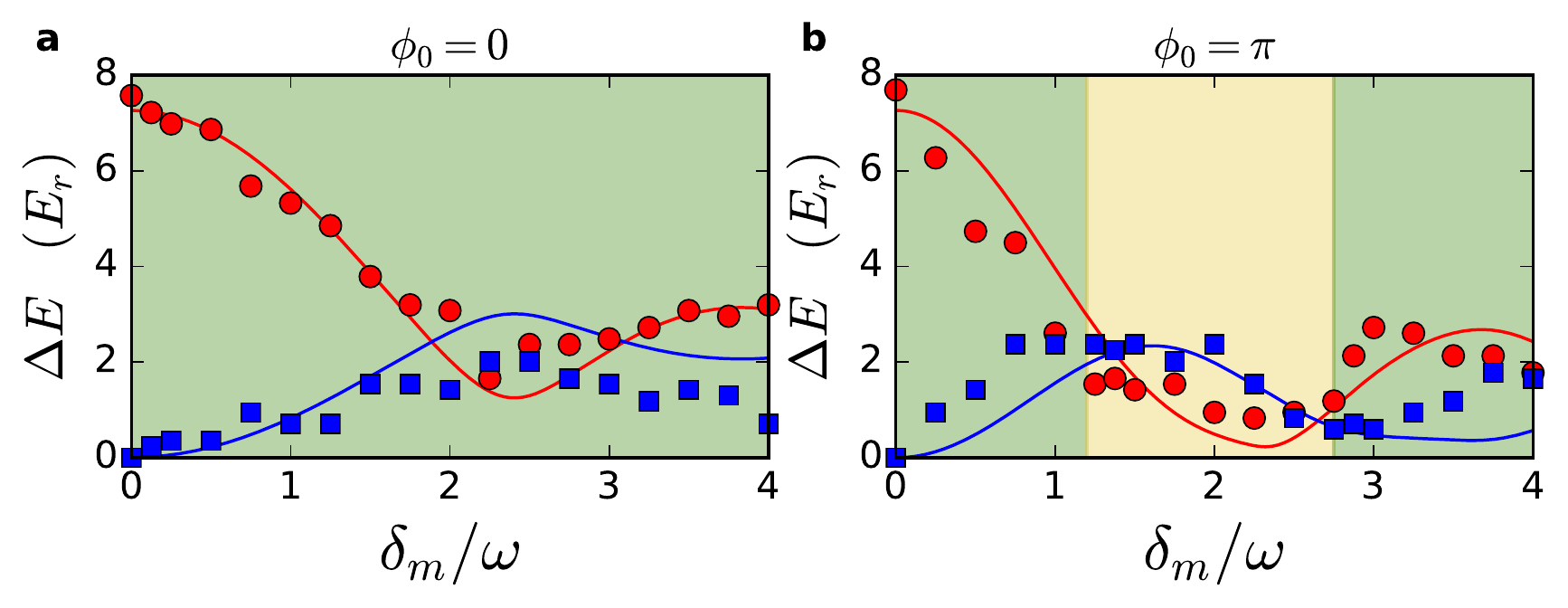}} 
\caption{\textbf{Effect of initial relative phase }$\protect\phi _{0}$. The
parameters and labels are the same as Fig.~\protect\ref{Fig3}(a) except that
$\protect\phi _{0}=0$ for (a) and $\protect\phi _{0}=\protect\pi $ for (b),
respectively.}
\label{fig:phase}
\end{figure}

\emph{Effects of relative phase}: In the presence of multiple modulations of
the system parameters, the relative phase between these modulations plays a
key role in the driven dynamics and the energy dispersions of corresponding
effective Hamiltonian are usually very different. A prominent example is the
comparison between circular and linear drivings of two components of a gauge
field where the former one breaks the time reversal symmetry and may lead to
the appearance of fascinating topological states while the latter one does
not~\cite{Wang2013, Jotzu2014}. Here the relative phase $\phi _{0}$ between
the modulation of the two detunings can dramatically change $\Omega
_{23}^{\prime }=J_{0}\left( 2\delta _{m}\sin (\phi _{0}/2)/\omega \right) $
and thus affect the sign of $\eta $ and the position of the Dirac point. In
Fig.~\ref{fig:phase}, we plot the band separations at the original Dirac
point $\mathbf{k}_{0}$, similar as Fig.~\ref{Fig3}(a), but with $\phi _{0}=0$
and $\phi _{0}=\pi $, respectively. When $\phi _{0}=0$, the Raman coupling $%
\Omega _{23}^{\prime }$ does not change sign during the modulation. The
simultaneous change of the other two Raman coupling strengths does not
reverse the sign of the parameter $\eta =\Omega _{12}^{\prime }\Omega
_{13}^{\prime }\Omega _{23}^{\prime }$, therefore the Dirac point always
exhibits at the lower two bands (Fig.~\ref{fig:phase}(a)). However, similar
to the case of $\phi _{0}=\pi /2$, $\eta $ changes sign for $\phi _{0}=\pi $%
, thus the Dirac point moves from the lower two bands to the upper two bands
and vice versa (Fig.~\ref{fig:phase}(b)).

{\Large Discussions}

The motion of atoms along the $z$ direction is decoupled from that in the $%
xy $ plane, therefore the rf spectroscopy only detects the band dispersion
in the $xy$ plane although the Fermi gas can be 3D. For a 2D (or a fixed $%
k_{z}$ plane in 3D) Fermi gas, a topological band gap at the Dirac point can
be opened by varying the polarizations of the Raman lasers, which induces an
imaginary part for the Raman coupling strength that corresponds to an
effective perpendicular Zeeman field. For example, in the recent experiment~%
\cite{Meng15}, an imaginary term $i\Gamma $ has been generated for the Raman
coupling $\Omega _{12}$. The exhibited energy gap $\Delta $ at the position
of the original Dirac point is found to be proportional to $\Gamma $ and the
Chern number of the two bands are given by $\pm \text{sgn}(\Gamma \Omega
_{13}\Omega _{23})$. In the presence of the same modulations that we
explored, the real and imaginary parts of the Raman coupling $\Omega _{12}$
change sign simultaneously. Consequently, the Chern number of the two gapped
bands are given by $\pm \text{sgn}(\eta )$ ($\mp \text{sgn}(\eta )$) if $%
\Gamma $ is of the same (opposite) sign with $\Omega _{12}$ before the
modulations are applied. This provides a useful guide to detect the
topological properties of the driven energy bands in the presence of an
energy gap. Such topological band gaps support the existence of exotic
Majorana fermions in 2D and Weyl fermions in 3D in the presence of pairing
interactions, while the periodic driving provides a knob of tuning
topological band regions, yielding Floquet Majorana or Weyl fermions.

In conclusion, we have experimentally directly observed the topology change
of the Floquet band structure in a periodically driven quantum system using
spin-injection rf spectroscopy. Our model system, periodically driven Fermi
gases with 2D SOC, provides an ideal platform for testing and understanding
rich Floquet physics and band engineering novel exotic quantum materials.

{\Large Methods}

\emph{Experimental setup}: The three spin states are selected within the $%
4^{2}S_{1/2}$ ground electronic manifold with $|1\rangle
=|F=7/2,m_{F}=1/2\rangle $, $|2\rangle =|9/2,1/2\rangle $ and $|3\rangle
=|9/2,3/2\rangle $, where $(F,m_{F})$ are the quantum numbers for hyperfine
spin states. The experiment starts with a Fermi gas of $N=2\times 10^{6}$ $%
^{40}$K atoms in a crossed 1064 nm optical dipole trap at $T/T_{F}\approx
0.3 $, where $T_{F}$ is the Fermi temperature defined by $%
T_{F}=(6N)^{1/3}\hbar \overline{\omega }/k_{B}$ with $\bar{\omega}\simeq
2\pi \times 80$ Hz labels the geometric trapping frequency. The fermionic
atoms are transferred into $|9/2,5/2\rangle $ as the initial state via a
rapid adiabatic passage induced by a rf field at $19.6$ G. Then a
homogeneous bias magnetic field along the $z$ axis (gravity direction) is
ramped to $B_{0}=121.4$ G by a pair of coils operating in the Helmholtz
configuration, splitting the $|3\rangle $ and $|2\rangle $ Zeeman states by $%
\sim $38.7 MHz and the $|1\rangle $ and $|2\rangle $ states by 1,293 MHz.
The large Zeeman splitting would isolate these three hyperfine spin states
from other ones in the Raman transitions. We choose the one-photon recoil
momentum $\hbar q_{r}$ and the recoil energy $E_{r}=\hbar
^{2}q_{r}^{2}/2m=h\times8.45$ kHz as the natural momentum and energy units.
Here $q_{r}=2\pi /\lambda $ and $\lambda $ is the wavelength of the Raman
lasers. Using the acoustic-optic modulators (AOM), the frequencies of the
Raman lasers 2 and 3 are modulated as $f_{2(0)}+\delta _{m2}\cos (\omega t)$
and $f_{3(0)}+\delta _{m3}\cos (\omega t+\phi _{0})$, respectively, yielding
the detuning modulations discussed above (see supplementary information \cite%
{supp}).

\emph{Effective Floquet Hamiltonian}: With the high-frequency modulation of
the Raman detunings, the motion of atoms in the $xy$ plane can be described
by an effective static Floquet Hamiltonian

\begin{equation}
H_{xy}^{eff}=\sum_{j=1}^{3}\left( \frac{\hbar ^{2}(\mathbf{k}-\mathbf{q}%
_{j})^{2}}{2m}+\delta _{j(0)}\right) |j\rangle \langle j|-\sum_{j^{\prime
}\neq j}\frac{\Omega _{jj^{\prime }}^{\prime }}{2}|j\rangle \langle
j^{\prime }|  \label{H_eff}
\end{equation}%
with the modified Raman coupling strengths $\Omega _{jj^{\prime }}^{\prime }$
due to the fast modulation. Here, $\hbar \mathbf{k}=(\hbar k_{x},\hbar
k_{y}) $ denotes the momentum of atoms projected on the $x-y$ plane, $\delta
_{2(0)} $ ($\delta _{3(0)}$) corresponds to the original two-photon Raman
detuning between Raman lasers $1$ and $2$ ($1$ and $3$) without the fast
modulation (i.e., $\delta _{1(0)}$ is chosen as 0). The wave vectors of
three lasers $\mathbf{q}_{1}=-q_{r}\hat{e}_{x}$, $\mathbf{q}_{2}=q_{r}\hat{e}%
_{y}$ and $\mathbf{q}_{3}=-q_{r}\hat{e}_{y}$.

\emph{Spin-injection spectroscopy}: The Raman lasers are derived from a
continuous-wave Ti-sapphire single frequency laser with the wavelength $%
\lambda =768.85$ nm which are ramped up linearly from zero to their final
intensity in 60 ms. Subsequently, a Gaussian shape pulse with 450 $\mu s$ of
the rf field is applied to drive atoms from $|9/2,5/2\rangle $ to the final
empty SOC state. Since the spin state $|9/2,5/2\rangle $ is coupled to the
state $|3\rangle $ via rf, spin injection rf spectroscopy will measure the
weight of the $|3\rangle $ state and obtain the energy dispersions with 2D
SOC. At last, the Raman lasers, the optical trap and the magnetic field are
switched off abruptly, and atoms freely expand for 12 ms in a magnetic field
gradient applied along the $x$ axis. Absorption image are taken along the $z$
direction. By counting the number of atoms in state $|3\rangle $ as a
function of the momentum and the rf frequency from the absorption image, the
energy band structure and the position of the Dirac point can be determined.

\textbf{Acknowledgements}: Useful discussions with L. Jiang are
acknowledged. This research is supported by the MOST (Grant No.
2016YFA0301600), NSFC (Grant No. 11234008, 11361161002, 11222430). C.Q. and
C.Z. are supported by by ARO (W911NF-17-1-0128), AFOSR (FA9550-16-1-0387),
and NSF (PHY-1505496).

\textbf{Author contributions}: L.H., P.P., D.L., Z.M., L.C., P.W., and J.Z.
performed experiments. C.Q., C.Z. and J.Z. developed the theory. C.Q., C.Z.,
and J.Z. contributed to the theoretical modelling and explanation of the
experimental data. C.Q., C.Z. and J.Z. wrote the paper. All authors
discussed the results and commented on the manuscript. J.Z. supervised the
project.

\textbf{Competing financial interests}

The authors declare no competing financial interests.

\newpage

\begin{widetext}

\textbf{Supplementary Information for ``Observation of Floquet band topology
change in driven ultracold Fermi gases"}

\section{Theoretical modelling}

The three Raman lasers propagate in the \textit{x-y} plane, thus the motion
of the atoms along the \textit{z} direction is decoupled from the internal
degrees of freedom. After the adiabatic elimination of the excited states,
the Hamiltonian for atoms can be written as $H=\hbar
^{2}k_{z}^{2}/(2m)+H_{xy}$ with
\begin{equation}
H_{xy}=\sum_{j=1}^{3}\left( \frac{\hbar ^{2}(\mathbf{k}-\mathbf{q}_{j})^{2}}{%
2m}+\delta _{j}\right) |j\rangle \langle j|-\sum_{j^{\prime }\neq j}\frac{%
\Omega _{jj^{\prime }}}{2}|j\rangle \langle j^{\prime }|.  \label{eq:Hxy}
\end{equation}%
Here, $\hbar \mathbf{k}=(\hbar k_{x},\hbar k_{y})$ denotes the momentum of
atoms projected on the $x-y$ plane, $\delta _{1}$ is set as zero (the energy
reference) for simplification and $\delta _{2}$ ($\delta _{3}$) corresponds
to the two-photon Raman detuning between Raman lasers $1$ and $2$ ($1$ and $%
3 $). The wavevectors of the three lasers are $\mathbf{q}_{j}$ with their
magnitudes given by $q_{r}=2\pi /\lambda $ where $\lambda $ is the
wavelength of the lasers. For the experimental configuration shown in
Fig.1(b) of the main text, the three vectors are given by $\mathbf{q}%
_{1}=-q_{r}\hat{e}_{x}$, $\mathbf{q}_{2}=q_{r}\hat{e}_{y}$ and $\mathbf{q}%
_{3}=-q_{r}\hat{e}_{y}$. The two-photon Raman coupling strength between the
between hyperfine states $j$ and $j^\prime$ is denoted by $\Omega
_{jj^{\prime }}$ which is a real number.

The Dirac point exhibited by the Hamiltonian Eq. (\ref{eq:Hxy}) is robust in
the sense that it moves in momentum space without gap opening as long as the
three Raman coupling strengths $\Omega _{jj^{\prime }}$ are real. The
position of the Dirac point can be manipulated by changing the magnitude or
the sign of the Raman coupling. Particularly, the Dirac point appears at the
crossing of the two lowest (highest) dressed bands if $\Omega _{12}\Omega
_{13}\Omega _{23}<0$ ($>0$). Below, we show that the periodic modulation of
the Raman detunings provides another powerful way for the manipulation of
the Dirac point and consequently alters the topology of the band structure.

Using the acoustic-optic modulators (AOM), the frequencies of the Raman
lasers 2 and 3 are modulated as $f_{2(0)}+\delta _{m2}\cos (\omega t)$ and $%
f_{3(0)}+\delta _{m3}\cos (\omega t+\phi _{0})$, respectively (see below
section B). Consequently, the two-photon Raman detunings become $\delta
_{2}=\delta _{2(0)}+\delta _{m2}\cos (\omega t)$ and $\delta _{3}=\delta
_{3(0)}+\delta _{m3}\cos (\omega t+\phi _{0})$ where $\phi _{0}$ is the
initial relative phase between the two modulations which could be tuned
arbitrarily in experiment. The overall phase of two modulations, which
amounts to a different launching time of the driving and may play a crucial
role for the micromotion of the atoms, is not important for effective model
description. However, the relative phase $\phi _{0}$ between the two
modulations always appear in the effective Hamiltonian and could change the
band structure dramatically.

The modulation frequency $\omega =2\pi /T=2\pi \times 100\text{ kHz}=\text{%
11.8}E_{r}$ is chosen to be much larger than other energy scales in Eq. (\ref%
{eq:Hxy}). For such a high-frequency driving, the system can be described by
the following time-independent effective Hamiltonian ($\hbar =1$)
\begin{equation}
H_{xy}^{\text{eff}}=\left(
\begin{array}{ccc}
\frac{(\mathbf{k}-\mathbf{q}_{1})^{2}}{2m} & -\frac{\Omega _{12}^{\prime }}{2%
} & -\frac{\Omega _{13}^{\prime }}{2} \\
-\frac{\Omega _{12}^{\prime }}{2} & \frac{(\mathbf{k}-\mathbf{q}_{2})^{2}}{2m%
}+\delta _{2(0)} & -\frac{\Omega _{23}^{\prime }}{2} \\
-\frac{\Omega _{13}^{\prime }}{2} & -\frac{\Omega _{23}^{\prime }}{2} &
\frac{(\mathbf{k}-\mathbf{q}_{3})^{2}}{2m}+\delta _{3(0)}%
\end{array}%
\right) ,  \label{static}
\end{equation}%
with the three effective Raman coupling strengths $\Omega _{jj^{\prime
}}^{\prime }$ are given by
\begin{eqnarray}
\Omega _{12}^{\prime } &=&\Omega _{12}\times \frac{1}{T}\int_{0}^{T}e^{-i%
\delta _{m2}\sin (\omega t)/\omega }dt=\Omega _{12}J_{0}\left( \frac{\delta
_{m2}}{\omega }\right) ,  \notag \\
\Omega _{13}^{\prime } &=&\Omega _{13}\times \frac{1}{T}\int_{0}^{T}e^{-i%
\delta _{m3}\sin (\omega t+\phi _{0})/\omega }dt=\Omega _{13}J_{0}\left(
\frac{\delta _{m3}}{\omega }\right) ,  \notag \\
\Omega _{23}^{\prime } &=&\Omega _{23}\times \frac{1}{T}\int_{0}^{T}e^{i%
\delta _{m2}\sin (\omega t)/\omega }e^{-i\delta _{m3}\sin (\omega t+\phi
_{0})/\omega }dt  \notag \\
&=&\Omega _{23}\bigg[J_{0}\left( \frac{\delta _{m2}}{\omega }\right)
J_{0}\left( \frac{\delta _{m3}}{\omega }\right) +2\sum_{n=1}^{\infty
}J_{n}\left( \frac{\delta _{m2}}{\omega }\right) \times  \notag \\
&&J_{n}\left( \frac{\delta _{m3}}{\omega }\right) \cos (n\phi _{0})\bigg]
\label{Ramanstrength}
\end{eqnarray}%
where $J_{n}(x)$ is the $n$-th order Bessel function of the first kind. Note
that the three effective Raman coupling strengths are still real and thus
the Dirac point is protected under the periodic modulation of the detunings.
In this work, we will consider the same modulation amplitudes for the two
detunings, i.e., $\delta _{m2}=\delta _{m3}=\delta _{m}$. In this situation,
the effective Raman coupling strength $\Omega _{23}^{\prime }$ can be
simplified to
\begin{eqnarray}
\Omega _{23}^{\prime } &=&\Omega _{23}\times \frac{1}{T}\int_{0}^{T}e^{-i2%
\delta _{m}\cos (\omega t+\phi _{0}/2)\sin (\phi _{0}/2)/\omega }dt  \notag
\\
&=&J_{0}\left( \frac{2\delta _{m}}{\omega }\sin (\phi _{0}/2)\right) .
\end{eqnarray}

Although the effective Raman couplings remain real numbers, because of the
nature of the Bessel functions, the signs of the effective Raman coupling
strengths $\Omega _{jj^{\prime }}^{\prime }$ and thus their product $\eta
=\Omega _{12}^{\prime }\Omega _{23}^{\prime }\Omega _{13}^{\prime }$ could
be reversed in the modulation (see Fig.1(c) of the main text). As a result,
the Dirac point may appear at the crossing of different dressed bands
depending on the laser parameters.

Below we show the derivation of the static effective Hamiltonian (\ref%
{static}).

\subsection{First method}

The time-dependent Hamiltonian is given by (we have taken $E_{r}={\hbar
^{2}k_{r}^{2}}/{2m}$ and $\hbar k_{r}$ as the units for the energy and
momentum),
\begin{equation}
H=\left(
\begin{array}{lll}
(\mathbf{k}-\mathbf{q}_{1})^{2} & -\frac{\Omega _{12}}{2} & -\frac{\Omega
_{13}}{2} \\
-\frac{\Omega _{12}}{2} & (\mathbf{k}-\mathbf{q}_{2})^{2}+\delta _{2}(t) & -%
\frac{\Omega _{23}}{2} \\
-\frac{\Omega _{13}}{2} & -\frac{\Omega _{23}}{2} & (\mathbf{k}-\mathbf{q}%
_{3})^{2}+\delta _{3}(t)%
\end{array}%
\right) .
\end{equation}%
The two detunings are modulated in the following way
\begin{eqnarray}
\delta _{2} &=&\delta _{2(0)}+\delta _{m2}\cos (\omega t+\alpha ), \\
\delta _{3} &=&\delta _{3(0)}+\delta _{m3}\cos (\omega t+\alpha +\phi _{0}),
\end{eqnarray}%
where $\alpha $ is the initial phase of the modulation and $\phi _{0}$ is
the relative phase between the two modulations.

To eliminate the time-dependence, one can apply a time-dependent unitary
transformation of the following form
\begin{equation}
U=\left(
\begin{array}{ccc}
1 & 0 & 0 \\
0 & e^{-i\frac{\delta _{m2}}{\omega }\sin (\omega t+\alpha )} & 0 \\
0 & 0 & e^{-i\frac{\delta _{m3}}{\omega }\sin (\omega t+\alpha +\phi _{0})}%
\end{array}%
\right) .  \label{eq:unitary}
\end{equation}%
The wave function is then transformed as $\tilde{\Psi}=U^{-1}\Psi $, i.e.,
\begin{equation}
\tilde{\Psi}=\left(
\begin{array}{c}
\tilde{\Psi}_{1} \\
\tilde{\Psi}_{2} \\
\tilde{\Psi}_{3}%
\end{array}%
\right) =\left(
\begin{array}{c}
\Psi _{1} \\
e^{i\frac{\delta _{m2}}{\omega }\sin (\omega t+\alpha )}\Psi _{2} \\
e^{i\frac{\delta _{m3}}{\omega }\sin (\omega t+\alpha +\phi _{0})}\Psi _{3}%
\end{array}%
\right)
\end{equation}%
The Hamiltonian is transformed as $\tilde{H}=U^{-1}HU-iU^{-1}\frac{\partial U%
}{\partial t}$, i.e.,
\begin{equation*}
\tilde{H}=\left(
\begin{array}{ccc}
(\mathbf{k}-\mathbf{q}_{1})^{2} & -\frac{\Omega _{12}}{2}e^{-i\frac{\delta
_{m2}}{\omega }\sin (\omega t+\alpha )} & -\frac{\Omega _{13}}{2}e^{-i\frac{%
\delta _{m3}}{\omega }\sin (\omega t+\alpha +\phi _{0})} \\
-\frac{\Omega _{12}}{2}e^{i\frac{\delta _{m2}}{\omega }\sin (\omega t+\alpha
)} & (\mathbf{k}-\mathbf{q}_{2})^{2}+\delta _{2(0)} & -\frac{\Omega _{23}}{2}%
e^{i\frac{\delta _{m2}}{\omega }\sin (\omega t+\alpha )}e^{-i\frac{\delta
_{m3}}{\omega }\sin (\omega t+\alpha +\phi _{0})} \\
-\frac{\Omega _{13}}{2}e^{i\frac{\delta _{m3}}{\omega }\sin (\omega t+\alpha
+\phi _{0})} & -\frac{\Omega _{23}}{2}e^{-i\frac{\delta _{m2}}{\omega }\sin
(\omega t+\alpha )}e^{i\frac{\delta _{m3}}{\omega }\sin (\omega t+\alpha
+\phi _{0})} & (\mathbf{k}-\mathbf{q}_{3})^{2}+\delta _{3(0)}\nonumber%
\end{array}%
\right) .
\end{equation*}%
The effective Hamiltonian is defined as the time-average of $\tilde{H}$ in
one period
\begin{equation}
H_{eff}=\frac{\omega }{2\pi }\int_{0}^{2\pi /\omega }\tilde{H}(t)dt.
\end{equation}%
Therefore the diagonal parts do not change while the non-diagonal parts will
be averaged out, yielding effective Raman couplings
\begin{eqnarray}
\Omega _{12}^{\prime } &=&\Omega _{12}\frac{\omega }{2\pi }\int_{0}^{2\pi
/\omega }e^{-i\frac{\delta _{m2}}{\omega }\sin (\omega t+\alpha )}dt=\frac{%
\Omega _{12}}{2\pi }\int_{0}^{2\pi }e^{-i\frac{\delta _{m2}}{\omega }\sin
(\tau +\alpha )}d\tau =\Omega _{12}J_{0}\left( \frac{\delta _{m2}}{\omega }%
\right) , \\
\Omega _{13}^{\prime } &=&\Omega _{13}\frac{\omega }{2\pi }\int_{0}^{2\pi
/\omega }e^{-i\frac{\delta _{m3}}{\omega }\sin (\omega t+\alpha +\phi
_{0})}dt=\frac{\Omega _{13}}{2\pi }\int_{0}^{2\pi }e^{-i\frac{\delta _{m3}}{%
\omega }\cos (\tau +\alpha +\phi _{0})}d\tau =\Omega _{13}J_{0}\left( \frac{%
\delta _{m3}}{\omega }\right) , \\
\Omega _{23}^{\prime } &=&\Omega _{23}\frac{\omega }{2\pi }\int_{0}^{2\pi
/\omega }e^{i\frac{\delta _{m2}}{\omega }\sin (\omega t+\alpha )}e^{-i\frac{%
\delta _{m3}}{\hbar }\sin (\omega t+\alpha +\phi _{0})}dt  \notag \\
&=&\Omega _{23}\bigg[J_{0}\left( \frac{\delta _{m2}}{\omega }\right)
J_{0}\left( \frac{\delta _{m3}}{\omega }\right) +2\sum_{n=1}^{\infty
}J_{n}\left( \frac{\delta _{m2}}{\omega }\right) \times J_{n}\left( \frac{%
\delta _{m3}}{\omega }\right) \cos (n\phi _{0})\bigg].
\end{eqnarray}

If $\delta _{m2}=\delta _{m3}$, then $\Omega _{23}^{\prime }$ can be
simplified to
\begin{eqnarray}
\Omega _{23}^{\prime } &=&\Omega _{23}\frac{1}{2\pi }\int_{0}^{2\pi }e^{i%
\frac{\delta _{m}}{\omega }[\sin (\tau +\alpha )-\sin (\tau +\alpha +\phi
_{0})]}d\tau =\frac{\Omega _{23}}{2\pi }\int_{0}^{2\pi }e^{-i\frac{2\delta
_{m}}{\omega }\sin \frac{\phi _{0}}{2}\cos (\tau +\alpha +\frac{\phi _{0}}{2}%
)}d\tau  \notag \\
&=&\Omega _{23}J_{0}\left( \frac{2\delta _{m}}{\omega }\sin (\phi
_{0}/2)\right) .
\end{eqnarray}%
Particularly, we find (i) $\Omega _{23}^{\prime }=\Omega _{23}$ for $\phi
_{0}=0$; (ii) $\Omega _{23}^{\prime }=\Omega _{23}J_{0}(\sqrt{2}\delta
_{m}/\omega )$ for $\phi _{0}=\pi /2$; (iii) $\Omega _{23}^{\prime }=\Omega
_{23}J_{0}(2\delta _{m}/\omega )$ for $\phi _{0}=\pi $.

\subsection{Second method}

The effective Hamiltonian can also be derived from the Goldman-Dalibard
method \cite{Goldman2014}. The equation is
\begin{equation}
H_{eff}=H_{0}+\frac{1}{\omega }\sum_{j=1}^{\infty }\frac{1}{j}%
[V^{(j)},V^{(-j)}]+\frac{1}{2\omega ^{2}}\sum_{j}^{\infty }\frac{1}{j^{2}}%
\left( [[V^{(j)},H_{0}],V^{(-j)}]+[[V^{(-j)},H_{0}],V^{(j)}]\right) +%
\mathcal{O}\left( \frac{1}{\omega ^{3}}\right) .
\end{equation}%
This methods becomes cumbersome for high order terms. But for high frequency
limit, we can safely keep the first several terms and find the approximate
effective Hamiltonian with good precision.

In the above expression, $H_{0}$ is the time-independent part of the
Hamiltonian and
\begin{equation}
V^{(+1)}=\left(
\begin{array}{ccc}
0 & 0 & 0 \\
0 & \frac{\delta _{m2}}{2}e^{i\alpha } & 0 \\
0 & 0 & \frac{\delta _{m3}}{2}e^{i(\alpha +\phi _{0})}%
\end{array}%
\right) ,\qquad V^{(-1)}=\left(
\begin{array}{ccc}
0 & 0 & 0 \\
0 & \frac{\delta _{m2}}{2}e^{-i\alpha } & 0 \\
0 & 0 & \frac{\delta _{m3}}{2}e^{-i(\alpha +\phi _{0})}%
\end{array}%
\right)
\end{equation}%
are positive and negative frequency parts of the periodic driving of $\delta
_{2}$ and $\delta _{3}$. Actually, the first order term vanishes and the
second term gives the effective Raman couplings
\begin{eqnarray}
\Omega _{12}^{\prime } &=&\Omega _{12}\left( 1-\frac{\delta _{m2}^{2}}{%
4\omega ^{2}}+\ldots \right) , \\
\Omega _{13}^{\prime } &=&\Omega _{13}\left( 1-\frac{\delta _{m3}^{2}}{%
4\omega ^{2}}+\ldots \right) , \\
\Omega _{23}^{\prime } &=&\Omega _{23}\left( 1-\frac{\delta _{m2}^{2}+\delta
_{m3}^{2}-2\delta _{m2}\delta _{m3}\cos \phi _{0}}{4\omega ^{2}}+\ldots
\right) .
\end{eqnarray}

If $\delta_{m2}=\delta_{m3}$, then $\Omega_{23}^\prime=\Omega_{23}\left[ 1-%
\frac{1}{4}\left(\frac{2\delta_m}{\omega}\sin\frac{\phi_0}{2}
\right)^2+\ldots \right]$. Using the series expansions for Bessel functions,
it is easy to find that the above expressions are the same (up to the order
of $1/\omega^2$) as that derived using the first method.

The so-called kick operator at time $t$ is given by
\begin{equation}
K(t)=\frac{1}{i\omega }\sum_{j=1}^{\infty }\frac{1}{j}\left(
V^{(j)}e^{ij\omega t}-V^{(-j)}e^{-ij\omega t}\right)
\end{equation}%
For our system, we have
\begin{equation}
K(t)=\left(
\begin{array}{ccc}
0 & 0 & 0 \\
0 & \frac{\delta _{m2}}{\omega }\sin (\omega t+\alpha ) & 0 \\
0 & 0 & \frac{\delta _{m3}}{\omega }\sin (\omega t+\alpha +\phi _{0})%
\end{array}%
\right),
\end{equation}%
which is diagonal. Therefore, the time-evolution operator for the original
Hamiltonian over a complete driving period takes the form
\begin{equation}
U(t=t_0+T, t_0)=e^{-iTH_{t_0}^{F}}=e^{-iK(t=t_0+T)}e^{-iTH_{eff}}e^{iK(t_0)}.
\end{equation}%
Here, the effective Hamiltonian $H_{eff}$ does not depend on the initial
time $t_0$. Its eigenvalues, i.e, the quasi-energies, determine the linear
phase evolution of the system. One should distinguish the effective
Hamitonian from $H_{t_0}^F$ which contains the information of the \textit{%
micromotion operator} $U_F(t)=e^{-iK(t)}$. It is easy to realize that $U_F(t)
$ is just the unitary operator (Eq.~(\ref{eq:unitary})) that we have
introduced in the previous method.

\section{Evolution of the chirality of the Dirac point in the modulation}

The low-energy effective Hamiltonian near the Dirac point can be written as
(see \cite{Meng15})
\begin{equation}
H_{eff}=\epsilon (\mathbf{p})\sigma _{0}+(-\alpha _{y}p_{y}+h_{x})\sigma
_{x}+(\beta _{y}p_{y}-\beta _{x}p_{x}+h_{z})\sigma _{z}
\end{equation}%
where $\alpha _{y}$, $h_{x}$, $\beta _{y}$, $\beta _{x}$, $h_{z}$ are
functions of the system parameters $\Omega _{12}$, $\Omega _{23}$, $\Omega
_{13}$, $\delta _{2}$, $\delta _{3}$, etc. The Dirac point appears at $%
p_{y}^{D}=h_{x}/\alpha _{y}$ and $p_{x}^{D}=(\beta
_{y}p_{y}^{D}+h_{z})/\beta _{x}$. By shifting the reference frame to this
point and redefining the momentum, one simplifies the above Hamiltonian to
\begin{equation}
H_{eff}=\epsilon (\mathbf{p})\sigma _{0}+(-\alpha _{y}p_{y})\sigma
_{x}+(\beta _{y}p_{y}-\beta _{x}p_{x})\sigma _{z}+\lambda \sigma _{y}
\end{equation}%
One finds~\cite{Xiao2010} that the chirality of the Dirac point is determined by
the coefficients $\alpha _{y}$ and $\beta _{x}$. Since $\text{sgn}(\alpha
_{y}\beta _{x})=\text{sgn}(\Omega _{13}\Omega _{23})$. Therefore, the
chirality reverses sign whenever one of the two Raman coupling $\Omega _{13}$
and $\Omega _{23}$ change sign.

\section{The configuration of Raman lasers}

The Raman lasers are derived from a continuous-wave Ti-sapphire single
frequency laser (M Squared lasers, SolsTiS) with the wavelength 768.85 nm as
shown in Fig. S1. The Raman laser 1 is sent through the two double-pass
acousto-optic modulators (AOM) (3200-124, Crystal Technology, Inc) driven by
two signal generators (N9310A, Agilent) and frequency shifted -212.975$%
\times $4 $\emph{MHz}$. The Raman laser 2 and 3 double-pass through two AOM
and are frequency-shifted +201.144$\times $2 and +220.531$\times $2 $\emph{%
MHz}$ respectively. In order to periodically drive the two-photon Raman
detuning, the Raman laser 2 and 3 are frequency modulated respectively with $%
f_{2(0)}+\delta _{m2}\cos (\omega t)$ and $f_{3(0)}+\delta _{m3}\cos (\omega
t+\phi _{0})$, in which a signal generator (AFG3252 Textronix) generates $%
\cos (\omega t)$ and $\cos (\omega t+\phi _{0})$ signal outputs
simultaneously to externally modulate the frequencies of two signal
generators (N5183A, Agilent) for Raman lasers 2 and 3. The modulation
frequency response of the frequency modulation of the signal generator
(N5183A, Agilent) may reach 3 $\emph{MHz}$ and the maximum deviation is
about 10 $\emph{MHz}$, which can satisfy the experimental requirement.

\begin{figure}[tbp]
\centerline{
\includegraphics[width=12cm]{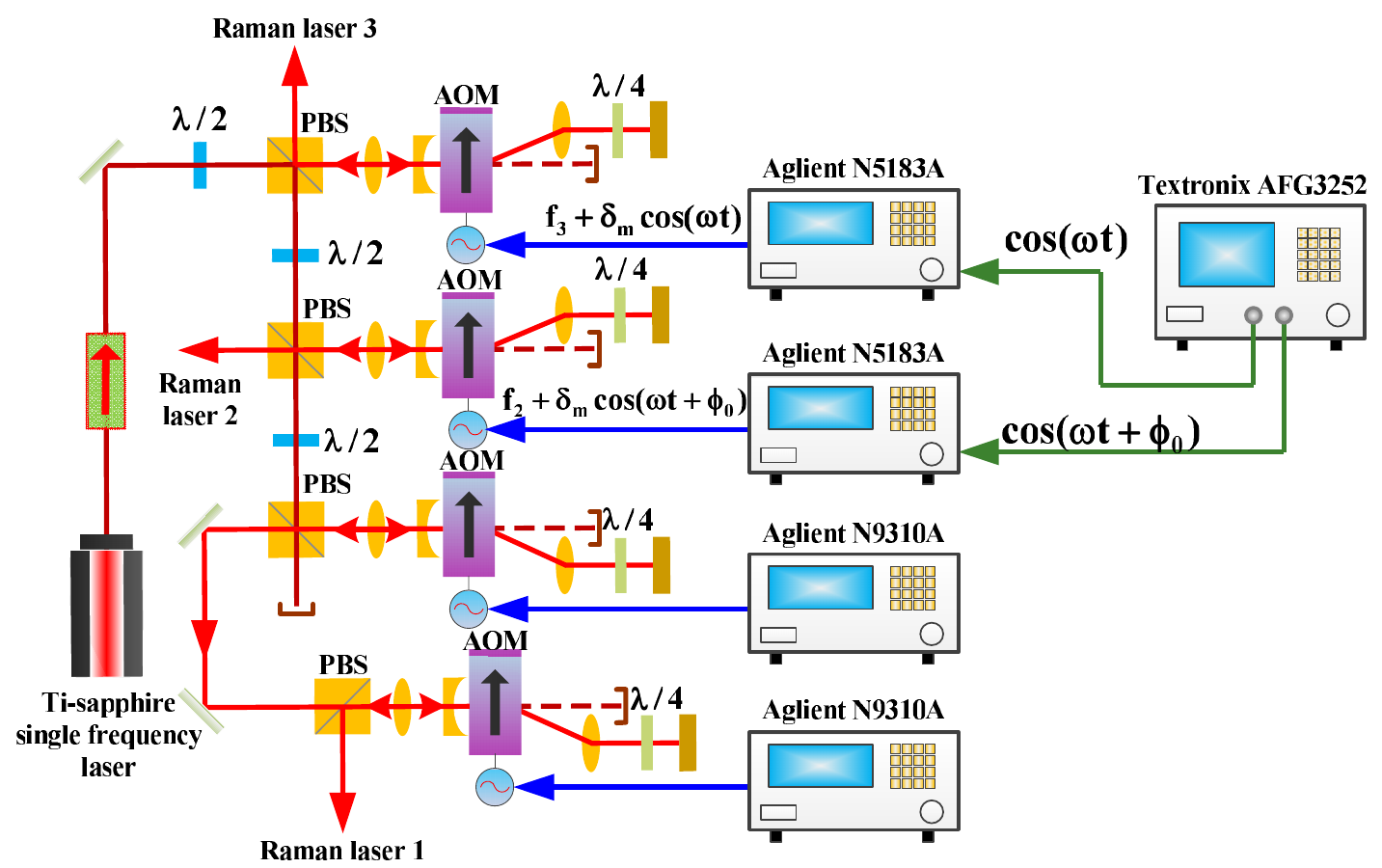}}
\caption{\textbf{Schematic of generating the Raman lasers.} $\protect\lambda %
/2$: half wave plate; $\protect\lambda /4$: quarter wave plate; PBS:
Polarized beam splitter; AOM: acousto-optic modulator. }
\label{Fig1}
\end{figure}

\end{widetext}

\end{document}